\documentclass[a4paper,12pt]{article}
\usepackage{graphicx}
\usepackage[small]{subfigure,epsfig}
\usepackage {amsmath} \usepackage{amssymb}
\usepackage{cite}
\usepackage[onehalfspacing]{setspace}
\usepackage{indentfirst}

\begin{document}

\title{On one of methods for finding exact solutions of nonlinear differential equations}


\author{Nikolai A. Kudryashov\footnote{nakudr@gmail.com}}

\date{Department of Applied Mathematics, National  Research Nuclear University MEPHI, 31 Kashirskoe
Shosse, 115409 Moscow, Russian Federation}

\maketitle


\begin{abstract}
One of old methods for finding exact solutions of nonlinear differential equations is considered. Modifications of the method are discussed. Application of the method is illustrated for finding exact solutions of the Fisher equation and nonlinear  ordinary differential equation of the seven order.  It is shown that the method is one of the most effective approaches for finding exact solutions of nonlinear differential equations.  Merits and demerits of the method are discussed.
\end{abstract}



\section{Introduction}

One of the first method for finding exact solutions of nonintegrable nonlinear partial differential equations was introduced in Ref. \cite{Kudr88} and applied in Ref. \cite{Kudr91}. However it seems to us that it was premature work.  At that time investigators did not use the symbolic calculations and the method \cite{Kudr88} did not came into notice.

Later we could observe appearance many other methods for finding exact solutions of nonlinear differential equations. Let us call here the following methods: the tanh - expansion method \cite{Parkes01, Malfliet01}, the simplest equation method \cite{Kudr90, Kudr05}, the Jacobi elliptic -  function method \cite{Parkes02}, the modified simplest equation method \cite{Vitanov01, Vitanov02, Vitanov03}, the cosh - function method \cite{Biswas, Biswas02}, the Exp - function method \cite{He01, He02}, the $G^{'}/G$ - expansion method \cite{Wang, Biswas01}. However some of the mentioned methods yield to the method by paper \cite{Kudr88}.

Currently we know four recent papers  \cite{Ryabov11, Kabir01, Kabir02, Gerpeel} of application of our method for finding exact solutions of nonlinear ordinary differential equations. However our opinion is that authors of these papers did not take into consideration all special particularities of our modified approach.

The aim of this paper is to present our approach again and give some examples of application of our method.

In fact we look for exact solution taking into account the following expression
\begin{equation}
  \begin{gathered}
  \label{Eq0}
  y(z)=\sum_{n=0}^{N}\,a_n\,Q^n,
      \end{gathered}
\end{equation}
where the function $Q$ takes the form
\begin{equation}
  \begin{gathered}
  \label{Eq0a}
 Q=\frac{1}{1+\mbox{e}^{z}}.
      \end{gathered}
\end{equation}

We note that the function $Q$ is solution of equation
\begin{equation}
  \begin{gathered}
  \label{Eq0b}
 Q_z=Q^2-Q.
      \end{gathered}
\end{equation}

This equation allows us to find derivatives $y_z$, $y_{zz}$ and so on. In the next section we present formulae for the six derivatives of solution $y(z)$.

We demonstrate application of our method for finding exact solutions of two nonlinear evolution equations.

One of these equations is the famous Fisher equation
\begin{equation}
  \begin{gathered}
  \label{Eq0c}
 u_t=u_{xx}+u\,(1-u)
      \end{gathered}
\end{equation}
and another equation of the seven order in the form
\begin{equation}
  \begin{gathered}
  \label{Eq0d}
 u_t+u^2\,u_x+\alpha\,u_{xxx}+\beta\,u_{xxxxx}-\gamma\,u_{xxxxxxx}=0.
      \end{gathered}
\end{equation}
We use  our method for obtaining exact solutions of these two
equations using the symbolic calculations.

\section{Algorithm of our method}

The algorithm of our method has six steps. They are the following.

\emph{The first step: reduction of nonlinear evolution equation to the ordinary differential equation.}

Let the partial differential equation in the polynomial form be given

\begin{equation}
  \begin{gathered}
  \label{Eq10a}
  E_1(u,u_t,u_x,u_{tt},u_{xx},\ldots)=0.
      \end{gathered}
\end{equation}

In the first step we use the reduction of the nonlinear partial differential equation to nonlinear ordinary differential equation taking the traveling wave solutions into account assuming
\[u(x,t)=y(z), \qquad z=k\,x+\omega\,t.\]

As result of this step we obtain the nonlinear ordinary differential equation
 with parameters $\omega$ and $k$  in  the form
\begin{equation}
  \begin{gathered}
  \label{Eq10b}
  E_2(y,\omega y_z,\,k y_z,\,\omega^2 y_{zz},\,k^2 y_{zz},\ldots)=0.
      \end{gathered}
\end{equation}

\emph{The second step: calculation of value $N$ in formula \eqref{Eq0} that is the pole order for the general solution of equation \eqref{Eq10b}.}

This value is determined taking the pole order of general solution for Eq.\eqref{Eq10b}. Substituting $y(z)=z^{-p}$, where $p>0$ into all monomials of Eq.\eqref{Eq10b} and comparing the two or more terms with smallest powers in equation we find the value for $N$.

We can use our method further if $N$ is the integer. In the case noninteger $N$ we have to use the transformation of solution $y(z)$. For example if we obtain the value for $N=\frac{1}{m}$, where $m$ is integer we can use the transformation for solution in the form $y(z)=w(z)^m$, where $w(z)$ is the new function.

\emph{The third step: substitution of derivatives for function $y(z)$ with respect to $z$ and the expression for $y(z$ into Eq.\eqref{Eq10b}.}

At this step  at the beginning we substitute derivatives of the function $y(z)$ with respect to $z$ starting with higher derivative into Eq.\eqref{Eq10b}. For a example, if we are going to find exact solutions of the nonlinear sixth order ordinary differential equation we substitute at the beginning the derivative of the sixth order
\begin{equation}
  \begin{gathered}
  \label{Eq11}
  y_{zzzzzz}=\sum_{n=0}^{N}\,a_n\,n\,Q^{n}\,(Q-1)\left[({n}^{5}+15\,{n}^{4}+85\,{n}^{3}+
  225\,{n}^{2}+274\,n+\right.
  \\
  \left.+120
 )\, Q^{5}-(5\,{n}^{5}+60\,{n}^{4}
+275\,{n}^{3}+600\,{n}^{2}+620\,n+240)Q^{4}+\right.
\\
+\left.(10\,{n}^{5}+90\,{n
}^{4}+ 315\,{n}^{3}+540\,{n}^{2}+ 455\,n+150)\,Q^{3}
-\right.\\
\left.-(10\,{n}^{5}+60\,{n}^{4}+145\,{n}^{3}+180
\,{n}^{2}+115\,n+30 )\, Q ^{2}+\right.\\
\left.
+(5\,{n}^{5}+15\,{n}^{4}+20\,{n}^{3}+15\,{n}^{2}+6\,n+1)\, Q-n^5
\right]
      \end{gathered}
\end{equation}
and then we have to substitute derivatives of fifth order, fourth order and so on using the following formulae
\begin{equation}
  \begin{gathered}
  \label{Eq10}
  y_{zzzzz}=\sum_{n=0}^{N}\,a_n\,n\,Q^{n}\,(Q-1)\left[(n^4+10n^3+35n^2+50n+24)\,Q^4
  +\right. \\
  \left.
  -(4n^4+30n^3+80n^2+90n+36)\,Q^3+(6\,n^4+30n^3+55n^2+\right. \\
  \left.+45n+14)\,Q^2
 -(4n^4+10n^3+
  10n^2+5n+1)\,Q+n^4\right],
      \end{gathered}
\end{equation}

\begin{equation}
  \begin{gathered}
  \label{Eq9}
  y_{zzzz}=\sum_{n=0}^{N}\,a_n\,n\,Q^{n}\,(Q-1)\left[(n^3+6n^2+11n+6)\,Q^3-\right. \\
  \left.
  -
  (3n^3+12n^2+15n+6)\,Q^2+(3n^3+6n^2+4n+1)\,Q-n^3\right],
      \end{gathered}
\end{equation}

\begin{equation}
  \begin{gathered}
  \label{Eq8}
  y_{zzz}=\sum_{n=0}^{N}\,a_n\,n\,Q^{n}\,(Q-1)\left[
  (n^2+3n+2)Q^2-\right. \\
  \left.
  -(2n^2+3n+1)\,Q+n^2\right],
      \end{gathered}
\end{equation}

\begin{equation}
  \begin{gathered}
  \label{Eq7}
  y_{zz}=\sum_{n=0}^{N}\,a_n\,n\,Q^{n}\,(Q-1)\left[(n+1)\,Q-n\right],
      \end{gathered}
\end{equation}

\begin{equation}
  \begin{gathered}
  \label{Eq6}
  y_z=\sum_{n=0}^{N}\,a_n\,n\,Q^{n}\,(Q-1).
      \end{gathered}
\end{equation}

At the end of this procedure we substitute the presentation of solution $y(z)$ in the form
\begin{equation}
  \begin{gathered}
  \label{Eq3}
  y=\sum_{n=0}^{N}\,a_n\,Q^n.
      \end{gathered}
\end{equation}

As result of the third step we obtain the equation that has the function  $Q$, coefficients $a_n$, where  $(n=0,1, \ldots, N)$ and parameters of $\omega$, $k$ of Eq.\eqref{Eq10b}.

\emph{Remark 2.1.} Some authors \cite{Ryabov11, Kabir01, Kabir02, Gerpeel} prefer to substitute $y(z)$ at the beginning. In this case we have to use derivatives function $Q(z)$ as at application of the simplest equation method for finding exact solutions.

\emph{Remark 2.2.} If we do not know how we can find the order of pole for the general solution we can use our algorithm as well but in this case we have to to take the large value of integer $N$.

\emph{Remark 2.3.} If Eq.\eqref{Eq10b} does not have some derivatives we do not substitute  expressions for them into equation.

\emph{Remark 2.4.} If we look for exact solutions of the differential equation with more then six order we have to calculate additional derivatives for function $y(z)$ using formulae  \eqref{Eq0} and \eqref{Eq0b}.

\emph{The fourth step: finding of algebraic equations for coefficients $a_n$ and for parameters $\omega$, $k$ and others of Eq.\eqref{Eq10b}.}

In the fourth step of our method we transform the problem of finding exact solution of nonlinear differential equation Eq.\eqref{Eq10b} into the problem of looking for solution of the system of algebraic equations.

Equating expressions at the different powers of $Q$ to zero we obtain the system of algebraic equations in the form
\begin{equation}
  \begin{gathered}
  \label{Eq3a}
      P_n(a_N,a_{N-1},\ldots, a_0,k,\omega, \ldots)=0,\qquad (n=0,\ldots, N).
      \end{gathered}
\end{equation}

\emph{The fifth step: solution of the system of algebraic equations.}

Solving the system of algebraic equations we obtain values of coefficients $a_N$, $a_{N-1}$, ..., $a_0$  and ralations for parameters of Eq.\eqref{Eq10b}. As result of solution we obtain exact solutions of Eq.\eqref{Eq10b} in the form \eqref{Eq3}.

\emph{The sixth step: the presentation of solution $y(z)$ of Eq.\eqref{Eq10b} in more convenient form and checking-up of solutions.}

Unfortunately there are many mistakes in finding exact solutions of
nonlinear differential equations and we have to give a good advice:
we need to test exact solutions of nonlinear differential equations.
The problem is that sometimes it is difficult to check-up exact
solutions of nonlinear differential equations. We believe that in
these cases we have to point out which solutions were checked up.

\section{Application of the method to the Fisher equation}

Consider the application of our method for looking exact solutions of the Fisher equation
\begin{equation}
  \begin{gathered}
  \label{Eq41}
 u_t=\delta\,u_{xx}+u\,(1-u).
      \end{gathered}
\end{equation}

1. As result of the first step we obtain the nonlinear ordinary differential equation in the form
\begin{equation}
  \begin{gathered}
  \label{Eq42}
\delta\,k^2\,y_{zz}-\omega\,y_{z}+y\,(1-y)=0.
      \end{gathered}
\end{equation}

2. In the second step we find $N=2$.

3. In the third step we substitute the second and the first derivatives  of function $y(z)$  into Eq.\eqref{Eq42}. In this case these derivatives can be written as

\begin{equation}
  \begin{gathered}
  \label{Eq43}
y_{zz}=Q \,(Q -1)  \left( 6\,{
a_2}\, Q ^{2}+2\,{a_1}\,Q
-4\,{a_2}\,Q -{a_1} \right),
      \end{gathered}
\end{equation}

\begin{equation}
  \begin{gathered}
  \label{Eq44}
y_{z}=Q\,(Q-1)\,(2\,a_2\,Q+a_1).
      \end{gathered}
\end{equation}

We have to substitute into Eq.\eqref{Eq42} the expression for $y(z)$ as well in the form
\begin{equation}
  \begin{gathered}
  \label{Eq45}
y=a_0+a_1\,Q+a_2\,Q^2.
      \end{gathered}
\end{equation}

As result of the third step we have the following equation
\begin{equation}
  \begin{gathered}
  \label{Eq46}
\left( 6\,\delta\,{k}^{2}{a_2}-{{a_2}}^{2} \right)  Q^4
+ \left(2
\,\delta\,{k}^{2}{ a_1}-10\,\delta\,{k}^{2}{a_2}-2\,\omega\,{a_2}-2\,{a_1}\,{a_2}
 \right)\,Q^3  + \\
 \\
 +\left( 2\,\omega\,{a_2}-3\,\delta\,
{k}^{2}{a_1}+4\,\delta\,{k}^{2}{a_2}-2\,{
a_0}\,{a_2}-\omega\,{a_1}-{{a_1}}^{2}+{a_2} \right)\,Q^2
+\\
\\
+ \left( \omega\,{a_1}+{
a_1}+\delta\,{k}^{2}{a_1}-2\,{a_0}\,{a_1} \right)\,Q  + {a_0}-{{a_0}}^{2}=0
      \end{gathered}
\end{equation}
4. Now we have to remember that $Q$ is the function and equation has solution when the
system is satisfied. It is possible when  the system of the following algebraic equations can be solved
 \begin{equation} \begin{gathered}
  \label{Eq47}
6\,\delta\,{k}^{2}{a_2}-{{a_2}}^{2} =0,\\
\\
2\,\delta\,{k}^{2}{ a_1}-10\,\delta\,{k}^{2}{a_2}-2\,\omega\,{a_2}-2\,{a_1}\,{a_2}=0,\\
\\
2\,\omega\,{a_2}-3\,\delta\,{k}^{2}{a_1}+4\,\delta\,{k}^{2}{a_2}-2\,{
a_0}\,{a_2}-\omega\,{a_1}-{{a_1}}^{2}+{a_2} =0,\\
\\
\omega\,{a_1}+{a_1}+\delta\,{k}^{2}{a_1}-2\,{a_0}\,{a_1}=0,\\
\\
{a_0}-{{a_0}}^{2}=0.
      \end{gathered}
\end{equation}

5. Solving the system of equations \eqref{Eq47} we have coefficients $a_2$, $a_1$ and $a_0$
in the form
 \begin{equation} \begin{gathered}
  \label{Eq48}a_2=6\,\delta\,{k}^{2},\quad a_1=-\frac{6\,\omega}{5}\,-6\,\delta\,{k}^{2},\\
  \\
  a_0=\frac12+\frac{3\,\omega}{5}\,+\frac{\delta\,{k}^{2}}{2}\,-{}\,{\frac {{\omega}
^{2}}{50\,\delta\,{k}^{2}}}
      \end{gathered}
\end{equation}
and the following values $\omega $ and $k$
 \begin{equation} \begin{gathered}
  \label{Eq49}\omega_{(1,2)}=\pm 5\,\delta\,k^2,
      \end{gathered}
\end{equation}

\begin{equation} \begin{gathered}
  \label{Eq49a}
  k_{(1,2)}=\pm\,\frac{1}{\sqrt{6\,\delta}}\quad k_{(3,4)}=\mp\,\frac{1}{\sqrt{-6\,\delta}}
      \end{gathered}
\end{equation}

Taking the formulae \eqref{Eq49} into account we obtain the  value for $\omega$
\[\omega_{(1)}=\frac56, \quad \omega_{(2)}=-\frac56.\]

6. Exact solutions of the Fisher equation takes the form

\begin{equation} \begin{gathered}
  \label{Eq50a}
 y_{(1,2)}(z)=\left(\frac{\exp{(z+z_0)}}{1+\exp{(z+z_0)}}\right)^2,\quad z=k_{(1,2)}\,x+\omega_{(1)}\,t
      \end{gathered}
\end{equation}

\begin{equation} \begin{gathered}
  \label{Eq50b}
 y_{(3,4)}(z)=1-\left(\frac{1}{1+\exp{(z+z_0)}}\right)^2,\quad z=k_{(3,4)}\,x+\omega_{(1)}\,t
      \end{gathered}
\end{equation}

\begin{equation} \begin{gathered}
  \label{Eq50c}
 y_{(5,6)}(z)=\left(\frac{1}{1+\exp{(z+z_0)}}\right)^2,\quad z=k_{(1,2)}\,x+\omega_{(2)}\,t
      \end{gathered}
\end{equation}

\begin{equation} \begin{gathered}
  \label{Eq50d}
 y_{(3,4)}(z)=\frac{1}{1+\exp{(z+z_0)}}\,\left(2-\frac{1}{1+\exp{(z+z_0)}}\right),\quad z=k_{(3,4)}\,x+\omega_{(2)}\,t
      \end{gathered}
\end{equation}

We check that all these solutions satisfy to the Fisher equation \eqref{Eq42}.

\section{Application of the method for the seven order nonlinear differential equation}

Let us demonstrate the application of our method for finding exact solutions of nonlinear partial differential equation of the seven order

\begin{equation}
  \begin{gathered}
  \label{Eq21}
 u_t+u^2\,u_x+\alpha\,u_{xxx}+\beta\,u_{xxxxx}-\gamma\,u_{xxxxxxx}=0.
      \end{gathered}
\end{equation}

Taking the travelling wave solutions in Eq.\eqref{Eq21} we have  the nonlinear ordinary differential equation after integration in the form
\begin{equation}
  \begin{gathered}
  \label{Eq22}
 \gamma\,k^7\,y_{zzzzzz}-\beta\,k^5\,y_{zzzz}+\alpha\,k^2\,y_{zz}+
\frac{k}{3}\,{y^3}+\omega\,y-C_1=0.
      \end{gathered}
\end{equation}

Note that if $y(z)$ is solution of Eq.\eqref{Eq22} then $-y(z)$ is also solution of this equation. It follows from the symmetry of Eq.\eqref{Eq22}. Substituting $y=a_0\,z^{-p}$ into all terms of equation and comparing terms we obtain $p=3$ and as consequence we have to take $N=3$ in solution \eqref{Eq3} .

As result of substitution of formulae \eqref{Eq3} and \eqref{Eq6} -- \eqref{Eq9} into Eq.\eqref{Eq22} we obtain the system of algebraic equations. Solving this system we obtain the following relations for coefficients $a_3$, $a_2$, $a_1$, $a_0$ and for parameters of Eq.\eqref{Eq22}

\begin{equation}
  \begin{gathered}
  \label{Eq23}
a_3=24\,\sqrt {105\,\gamma\,}\,{k}^{3},\quad a_2=-36\,\sqrt {105\,\gamma}\,{k}^{3}, \\
 \\
 a_1=12\,\sqrt {
105\,\gamma\,}\,{k}^{3}-\,{\frac {12\,\beta\,k\sqrt {105}}{83\,\sqrt {\gamma}}},\quad a_0=\,{\frac {6\,\beta\,k\sqrt {105}}{83\,\sqrt {\gamma}}}
      \end{gathered}
\end{equation}
In formula \eqref{Eq23} the coefficient $\alpha_3$ has signs plus and minus but we use only one sign.

\begin{equation}
  \begin{gathered}
  \label{Eq24a}
\alpha= 21\,\gamma\,{k}^{4}+{\frac {210}{83}}\,\beta\,{k}^{2}-{\frac {1177\,{\beta}^{2}}{6889\,\gamma}},
      \end{gathered}
\end{equation}

\begin{equation}
  \begin{gathered}
  \label{Eq24}
\omega=\,{\frac {{1343\,\beta}
^{3}\,k}{{571787\,\gamma}^{2}}}-\,{\frac {2310\,{k}^{3}{\beta}^{2}}{6889\,\gamma}}-{\frac {273}{83
}}\,\beta\,{k}^{5}-20\,\gamma\,{k}^{7}
      \end{gathered}
\end{equation}

\begin{equation}
  \begin{gathered}
  \label{Eq25}
  k_{(1,2)}=\pm\,\sqrt {\frac {{\beta}}{{83\,\gamma}}},\quad k_{(3,4)}=\pm\,{\frac {1}{1660}}\,{\frac {\sqrt {830}\sqrt {\gamma\,\beta\, \left( -293+3\,i
\sqrt {2399} \right) }}{\gamma}},\\
\\
k_{(5,6)}=\pm\,{\frac {1}{1660}}\,{\frac {\sqrt {-830\,\gamma\,\beta\, \left( 293+3\,i\sqrt
{2399} \right) }}{\gamma}}.
       \end{gathered}
\end{equation}

\begin{equation}
  \begin{gathered}
  \label{Eq26}
  C_1={\frac {}{}}\,{\frac {6300{k}^{4}\sqrt {105}{\beta}^{3}}{{571787\gamma}^{3/
2}}}+{\frac {1638{k}^{6}{\beta}^{2}\sqrt {105}}{
6889\sqrt {\gamma}}}+\\
\\
+{\frac {120}{83}}\,\beta\,{k}^{8}\sqrt {105\gamma}-{
\frac {}{}}\,{\frac {8058{k}^{2}\sqrt {105}{\beta}^{4}}{47458321{\gamma}^{5
/2}}}.
      \end{gathered}
\end{equation}

So we have exact solutions of Eq.\eqref{Eq22} in the form

\begin{equation}
  \begin{gathered}
  \label{Eq27}
y(z)={\frac {6\,\beta\,k\sqrt {105}}{83\,\sqrt {\gamma}}}\,-\left(12\,\sqrt {
105\,\gamma\,}\,{k}^{3}-\,{\frac {12\,\beta\,k\sqrt {105}}{83\,\sqrt {\gamma}}}\right)\,Q(z)-\\
\\
- 36\,\sqrt {105\,\gamma}{k}^{3}\,Q(z)^2+24\,\sqrt {105\,\gamma\,}{k}^{3}\,Q(z)^3,\quad Q(z)=\frac{1}{1+\exp{(z+z_0)}},
      \end{gathered}
\end{equation}
where values $\omega$ and $k$ are determined by formulae \eqref{Eq24} and \eqref{Eq25}. These solutions exist at relation for parameter $\alpha$ that is found from expression \eqref{Eq24a}. Exact solutions \eqref{Eq27} at $k=k_{(1,2)}$ were checked-up by us.

\section{Merits and demerits of the method}

Let us discuss merits and demerits of our method. The first merit is that the method is simple in its application. The second merit of the method is that we can use the united formulae for all nonlinear differential equations in the polynomial form. What is more we can not calculate the value $N$ for solution. However in this case it is better to take the big value of integer $N$.

The third merit of our approach is that this one allows us to obtain all solitary wave solutions and all one periodic solutions when we get the expansion of the general solution of nonlinear differential equation in the Laurent series \cite{Kudr10}.

The fourth merit of our approach is we can have solution taking the different functions into account. One can see that the function $Q(z)$ can have different forms.
For a example we obtain that using the equality
\begin{equation}
  \begin{gathered}
  \label{Eq31}
Q(z)=\frac{1}{1+\mbox{e}^{z}}=\frac{1}{1+\mbox{e}^{z}}-\frac12+\frac12=\frac12
\left[1-\tanh{\left(\frac{z}{2}\right)}\right].
      \end{gathered}
\end{equation}

We see that exact solution for nonlinear differential equation can be presented using the hyperbolic tangents.

Taking the autonomous form of nonlinear differential equation we can use the additional constant $z_0$ in function $Q(z)$. In this case we have to write the function $Q(z)$ in the form
\begin{equation}
  \begin{gathered}
  \label{Eq32}
Q(z)=\frac{1}{1+\mbox{e}^{(z+z_0)}}=\frac12
\left[1-\tanh{\left(\frac{z}{2}+\frac{z_0}{2}\right)}\right]=\\
\\
=\frac{1}{2}\left[1-
\coth\left(\frac{z}{2}+\frac{z_1}{2}\right)\right],
      \end{gathered}
\end{equation}
where $z_0$ is arbitrary constant and $z_1$ is the constant as well that is
\[z_0=\frac{i\,\pi}{2}+z_1.\]
We see that using $Q(z)$ we can present solution via the $\coth(z)$ functions.
One can suggest many other trigonometric functions for description of solutions of nonlinear differential equations that can be obtained from function $Q(z)$.

Assuming
\[Q(z)=\frac{G^{'}}{G}\]
we obtain from formula \eqref{Eq3} expressions for the $\frac{G^{'}}{G}$ -- expansion method.

It is important to note that our method gives all results that can be found by  the Exp -- function method. It is known (see,for a example Refs. \cite{Kudr09a, Kudr09b, Kudr09c}) how many mistakes gave the Exp-function method. This method does not allow us to look for exact solutions for higher order nonlinear differential equations.

Let us shortly discuss demerits of our approach.
The first demerit of our approach is that this one does not allow us to find all solutions for the equation for cases when there are two and more brunches of expansion for the general solution in  the Laurent series. However we have to note most of other methods do not allow to find such solutions as well. In these cases we have to use more difficult methods that were introduced recently in works \cite{Kudr10a, Kudr10b, Kudr11}.

The second demerit of our method is that the approach does not allow us to look for two periodic solutions. For this aim we have to use more difficult methods that were developed in our recent works \cite{Kudr10a, Kudr10b, Kudr11}.

\section{Acknowledgements}

Author is grateful to M.M. Kabir for sending papers \cite{Kabir01, Kabir02}.
This research was supported by Federal Target Programm Research and Scientific Pedagogical Personnnel of Innovation in Russian Federation on 2009 -- 2013.

\end{document}